\documentclass[12pt]{iopart}
\usepackage{graphicx}
\usepackage{iopams}
\usepackage{amsfonts}
\newcommand{\vnabla}{{\mbox{\boldmath$\nabla$}}}

\newcommand{\vJ}{{\mbox{\boldmath$J$}}}
\newcommand{\vr}{{\mbox{\boldmath$r$}}}

\newcommand{\vk}{{\mbox{\boldmath$k$}}}
\newcommand{\vv}{{\mbox{\boldmath$v$}}}

\newcommand{\vsk}{{\small \mbox{\boldmath$k$}}}

\newcommand{\vg}{\mbox{\boldmath$g$}}

\newcommand{\hvg}{\hat{\mbox{\boldmath$g$}}}

\newcommand{\vsig}{\mbox{\boldmath$\sigma$}}
\newcommand{\vd}{\mbox{\boldmath$d$}}
\newcommand{\mhx}{\hat{\mbox{\boldmath$x$}}}
\newcommand{\mhy}{\hat{\mbox{\boldmath$y$}}}
\newcommand{\mhz}{\hat{\mbox{\boldmath$z$}}}

\newcommand{\vH}{\mbox{\boldmath$H$}}
\newcommand{\vS}{\mbox{\boldmath$S$}}
\begin{document}
%
\title[]{Spin susceptibility in superconductors without inversion symmetry}

\author{P. A. Frigeri$^1$, D.F. Agterberg$^2$ and M. Sigrist$^1$
}

\address{$^1$ Theoretische Physik, ETH-H\"onggerberg, 8093 Z\"urich,
  Switzerland \\
  $^2$ Department of Physics, University of Wisconsin-Milwaukee, Milwaukee, Wisconsin 53201, USA}

\begin{abstract}
In materials without spatial inversion symmetry the spin
degeneracy of the conduction electrons can be lifted by an
antisymmetric spin-orbit coupling. We discuss the influence of
this spin-orbit coupling on the spin susceptibility of such
superconductors, with a particular emphasis on the recently
discovered heavy Fermion superconductor CePt$_3$Si. We find that,
for this compound (with tetragonal crystal symmetry,) irrespective
of the pairing symmetry, the stable superconducting phases would
give a very weak change of the spin susceptibility for fields
along the c-axis and an intermediate reduction for fields in the
basal plane. We also comment on the consequences for the
paramagnetic limiting in this material.
\end{abstract}

\maketitle

\section{Introduction}

Our understanding of conventional superconductors is based on the
BCS theory of Cooper pairing induced by electron-phonon
interaction. The electrons pair up in the superconducting phase
with the most symmetric pair wave function possible, i.e. in the
s-wave orbital and spin-singlet channel. According to Anderson
(1959) this type of pairing requires time reversal symmetry,
because  the paired electrons need to be in degenerate states of
opposite momentum and opposite spin \cite{Anderson1959}.  This
theorem guarantees that the conventional superconducting phase
remains stable even, if a sample is dirty as long as time reversal
symmetry is present and allows us to construct degenerate electron
pairs with vanishing total momentum.

The situation becomes more complex for unconventional
superconductivity with Cooper pairs of lower symmetry. Impurities
of all kind are detrimental to such pairing due to the momentum
averaging effect of random potential scattering. For clean samples
new symmetry criteria have been formulated by Anderson (1984)
\cite{Anderson1984}. While spin-singlet pairing only requires time
reversal symmetry, spin-triplet pairing also requires inversion
symmetry. These symmetries are present in most materials and no
discussion on this issue is usually required.

Motivated by the recent discovery of the heavy Fermion
superconductor CePt$_3$Si which has no inversion center, we will
discuss the magnetic properties of the superconducting phase
\cite{Bauer2004} While the absence of an inversion center does not
imply immediately unusual behavior of the superconducting phase
special interest in this case has arisen with the observation that
the upper critical field dramatically exceeds the paramagnetic
limit. For heavy Fermion superconductors the coherence length is
generally rather short, roughly 100 \AA, giving rise to large
orbital depairing fields. Thus, if there were a paramagnetic
limiting effect, then it would be likely be observed; as  for
example in CeCoIn$_5$ \cite{C115}. CePt$_3$Si has a critical
temperature of $ T_c =0.75 K $ and a zero-temperature extrapolated
upper critical field of 5 Tesla. The estimated paramagnetic limit
is a factor 5 smaller than $ H_{c2} $ ($ H_p \approx \Delta /
\sqrt{2} \mu_B \approx 1 $ Tesla) \cite{Bauer2004}. Paramagnetic
limiting is expected for spin-singlet pairing, as the Zeeman
coupling to the spins would break up the Cooper pairs. Since
Cooper pairs in spin-triplet configurations would not be destroyed
by spin polarization, it might be concluded that spin-triplet
pairing is realized in CePt$_3$Si. However, the absence of an
inversion center appears to be an obstacle for spin-triplet
pairing.

It has been shown that broken inversion symmetry is not
indiscriminately destructive for the spin-triplet pairing, and,
additionally, it softens the effect of the paramagnetic depairing
for spin-singlet pairs \cite{Frigeri2004}. For many trpes of
crystal lattices, the absence of an inversion center introduces an
antisymmetric spin-orbit coupling (SOC), analogous to the
well-known Rashba type of spin-orbit coupling. Naturally this
influences the Cooper pairing through the modification of the band
structure as shown by various groups
\cite{Frigeri2004,Gorkov2001,Samokhin}.

We would like to discuss here the problem of the spin
susceptibility of the superconducting phase, since this is also
directly connected with the issue of paramagnetic limiting. The
spin susceptibility measured by means of the NMR-Knight shift in
the superconducting phase is often used to distinguish between
spin-singlet and spin-triplet pairing. We will show here that this
kind of discrimination between the two types of states is not
generally possible anymore. Rigorously speaking it is, of course,
not possible to separate spin-singlet and triplet in the presence
of SOC. We will consider here weak SOC which we can turn on
adiabatically to follow the evolution of the originally
well-defined singlet and triplet pairing states. Our basic result
is that the spin susceptibility of the spin-singlet states
gradually approaches the behavior of the spin-triplet state which
survives the presence of the spin-orbit coupling. Moreover we can
predict that the spin susceptibility for fields along the z-axis
of the tetragonal crystal lattice of CePt$_3$Si would be less
suppressed than for fields along the basal plane.

%
\section{The basic Model and normal state Greens function}
%
For the following discussion we will use the model introduced by Frigeri et al. \cite{Frigeri2004} for
CePt$_3$Si,
which has the following single-particle Hamiltonian:

\begin{equation}
{\cal H}_0 = \sum_{\vsk,s,s'} \left[ \xi_{\vsk} \sigma_0 + \alpha
  \vg_{\vsk} \cdot \vsig \right]_{ss'} c_{\vsk s}^{\dag} c_{\vsk s'}
\label{eq-1}
\end{equation}
where $ c_{\vsk s}^{\dag} $ ($ c_{\vsk s} $) creates (annihilates)
an electron with  momentum $ \vk $  and spin $s$. The band energy
$ \xi_{\vsk} = \epsilon_{\vsk} - \mu $ is measured relative to the
chemical potential $ \mu $ and $ \alpha \vg_{\vsk} \cdot \vsig $
introduces the antisymmetric spin-orbit coupling with $ \alpha $
as a coupling constant (we set $ \langle \vg_{\vsk}^2
\rangle_{\vsk} = 1 $ where $\langle \rangle$ denotes the average
over the Fermi surface).

We give here a brief discussion on the origin and some basic
properties of the antisymmetric SOC. In a crystal lattice the
electrons move in a periodic potential $ U(\vr) $. In the absence
of inversion symmetry there is no symmetry point in the unit cell
relative to which $ U(\vr) = U(-\vr) $ is satisfied. This also
implies that the Bloch function does not have the property that $
u_{\vsk}(\vr) = u_{-\vsk} (- \vr) $. Ignoring for the moment
relativistic effects, the potential yields the following
contribution to the single-particle Hamiltonian:
\begin{eqnarray}
{\cal H}_p & = & \sum_{\vsk,s} \int_{u.c.} d^3r \, u^*_{\vsk}(\vr) U(\vr) u_{\vsk}(\vr) \, c_{\vsk s}^{\dag} c_{\vsk s}
\nonumber \\
&=&  \sum_{\vsk,s} \int_{u.c.} d^3r \, u_{-\vsk}(\vr) U(\vr) u_{\vk}(\vr) \, c_{\vsk s}^{\dag} c_{\vsk s}
= \sum_{\vsk,s} \tilde{U}(\vsk) \, c_{\vsk s}^{\dag} c_{\vsk s} \;.
\end{eqnarray}
The integral runs over the unit cell of the lattice. The resulting potential $ \tilde{U}(\vk) $ is even
in $\vk$, i.e. $ \tilde{U}(\vk) = \tilde{U}(-\vk) $. On this (non-relativistic) level the lack of inversion symmetry
does not affect the band structure which remains symmetric under the operation $ \vk \to - \vk $ due to time reversal
symmetry, i.e. the fact that $ u^*_{\vsk} (\vr) = u_{-\vsk} (\vr) $.

Now we include the SOC. The symmetric SOC of each ion couples different atomic orbitals and
requires a multi-orbital description. In this way the spins would be converted into pseudospins
which can be handled formally in the same way as the original spins.  For our purpose the
most important influence of this would be an anisotropic $g$-tensor which could be straightforwardly
introduced into the following discussion and would affect the susceptibility of the normal and superconducting phase in the same way. For the sake of simplicity, however, we will keep here the
$g$-tensor isotropic. Moreover we ignore the effect of spin-orbit coupling on the pairing interaction.

For a lattice without inversion symmetry SOC already appears on
the level of a single-band model yielding the second term in Eq.(\ref{eq-1}) \cite{Dresselhaus1955}. The vector function
$ \alpha \vg_{\vsk} $ is derived from the relativistic correction $
\frac{e}{2 mc^2} [\vv \times \vnabla_{\vr} U(\vr) ] \cdot \vS $, which yields

\begin{equation}
\alpha \vg_{\vsk} =- \frac{e}{2mc^2} \int_{u.c.} d^3r \left\{ \vJ_{\vsk}(\vr) \times
\vnabla_{\vr} U(\vr) \right\}
\end{equation}
with
\begin{equation}
\vJ_{\vsk} (\vr) = \frac{\hbar}{2 mi} \left[ u^*_{\vsk} (\vr) (i\vk + \vnabla_{\vr} ) u_{\vsk} (\vr) + u_{\vsk} (\vr) (i\vk - \vnabla_{\vr} ) u^*_{\vsk} (\vr) \right] \;.
\end{equation}
It is easy to verify that $ \alpha \vg_{\vsk} = 0 $, if $ U(\vr) =
U(-\vr) $ and $ u_{\vsk} (-\vr) = u_{-\vsk} (\vr) $. In the
absence of inversion symmetry, however, $ \vg_{\vsk} $ is finite
and satisfies $ \vg_{\vsk} = - \vg_{-\vsk} $, since $
\vJ_{-\vsk}(\vr) = - \vJ_{\vsk} (\vr) $ (note if  $\vg_{\vsk} =
\vg_{-\vsk}$ this implies time reversal symmetry is broken). With
these properties it is now clear that the Hamiltonian $ {\cal H}_0
$ is invariant under time reversal $ {\cal T} $ but not under
inversion operation $ {\cal I} $, because

\begin{eqnarray}
{\cal I } \{ \alpha   \vg_{\vsk} \cdot \hat{\vsig} \} {\cal I}^{-1} = - \alpha
\vg_{\vsk} \cdot \hat{\vsig}  \qquad \mbox{and} \qquad {\cal T } \{ \alpha
\vg_{\vsk} \cdot \hat{\vsig} \} {\cal T}^{-1} =  \alpha \vg_{\vsk}
\cdot \hat{\vsig}  \nonumber.
\end{eqnarray}

The SOC yields a modified band structure. We parameterize the
Green's function by
\begin{eqnarray}
\hat{G}_0(\vk, i \omega_n) = G^{0}_{+} (\vk, i \omega_n) \hat{\sigma}_{0} +
  (\hvg_{\vsk} \cdot  \hat{\vsig}) G^{0}_{-}(\vk, i \omega_n)
\end{eqnarray}
where
\begin{eqnarray}
G_{\pm} (\vk, i \omega_n) & = &\frac{1}{2} \left[ \frac{1}{i \omega_n -
    \xi_{\vsk} - \alpha |\vg_{\vsk}|} \pm \frac{1}{i \omega_n -
\xi_{\vsk} + \alpha | \vg_{\vsk}|} \right]
\end{eqnarray}
and  $ \hvg_{\vsk} = \vg_{\vsk} / | \vg_{\vsk} | $ ($ | \vg | =
\sqrt{\vg^2} $). The band splits into two spin dependent parts
with energies $ E_{\vsk,\pm} = \xi_{\vsk} \pm \alpha | \vg_{\vsk}|
$. The spinor is twisted on the two bands in a way that is
described by the antisymmetric part of the Green's function, $ (\hvg_{\vsk} \cdot  \hat{\vsig})
G^0_- (\vk, i \omega_n) $.

\section{Superconducting phase}

Now we turn to the superconducting phase and introduce the general pairing interaction
\begin{eqnarray}
{\cal H}_{pair} = \frac{1}{2} \sum_{\vsk, \vsk, s_i}  V_{s_1,\cdots,
                     s_4}(\vk,\vk') \; c^{\dag}_{\vsk,s_1}
                     c^{\dag}_{-\vsk,s_2} c_{-\vsk',s_3} c_{\vsk',s_4}
\end{eqnarray}
The interaction satisfies the relations $V_{\alpha, \beta ;
  \gamma \delta} (\vk,\vk')=-V_{ \beta, \alpha; \gamma,
  \delta}(-\vk,\vk')=-V_{\alpha, \beta ; \delta, \gamma}(\vk,-\vk')=V_{\delta,
  \gamma; \beta, \alpha }(\vk',\vk)$ to assure that the Fermion sign and
the time reversal symmetry ($\cal T$) are preserved.

For the following calculations it is advantageous to discuss the superconducting phase
by means of Green's functions
\begin{eqnarray}
        \label{green_function}
        G_{\lambda \mu}(\vk, \tau)&=&- \langle T_{\tau} \{
        c_{\vsk,\lambda}(\tau) c^{\dag}_{\vsk,\mu}(0)\} \rangle \nonumber \\
        F_{\lambda \mu}(\vk, \tau )&=& \langle T_{\tau} \{
        c_{\vsk,\lambda}(\tau) c_{-\vsk,\mu}(0)\} \rangle \nonumber \\
        F^{\dag}_{\lambda \mu}(\vsk, \tau )&=& \langle T_{\tau} \{
        c^{\dag}_{-\vsk,\lambda}(\tau) c^{\dag}_{\vsk,\mu}(0) \}
        \rangle, \nonumber
\end{eqnarray}
where the operators $c_{\vsk,\lambda}(\tau)$ and
$c^{\dag}_{\vsk,\lambda}(\tau)$  are expressed in the Heisenberg
representation.  These Green's functions have to satisfy
Gor'kov equations of the following form
 \begin{eqnarray}
        \label{Green1}
        \left[\hat{G}^{-1}_0(\vk,i\omega_n)+\hat{\Delta}(\vk)\hat{G}^{\top}_0(
        -\vk,-i\omega_n)\hat{\Delta}^{\dag}(\vk)\right]\hat{G}(\vk,i\omega_n)=
        \hat{\sigma}_0. \\
        \hat{F}(\vk,i\omega_n)=\hat{G}_{0}(\vk,i\omega_n)\hat{\Delta}(\vk)
        \hat{G}^{\top}(-\vk,-i\omega_n). \label{Green2} \\
        \hat{F}^{\dag}(\vk,i\omega_n)=\hat{G}^{\top}_0( -\vk,-i\omega_n)
        \hat{\Delta}^{\dag}(\vk)\hat{G}(\vk,i\omega_n) \label{Green3}
\end{eqnarray}
where $\hat{\Delta}^{\dag}(\vk)$ is the gap function defining
the order-parameter of the superconducting state and is $2 \times 2$-matrix
in spin space.

In general the gap function has a singlet $\psi(\vk)$ and a
triplet component $\vd(\vk)$, i.e. $\hat{\Delta}^{\dag}(\vk)=i \{
\psi(\vk)+\vd(\vk)\cdot \hat{\vsig} \} \hat{\sigma}_{y}$,  where
$\psi(\vk)=\psi(-\vk)$ is even and $\vd(\vk)=-\vd(-\vk)$ is odd. For a finite
$\alpha$, the gap equations for the spin-singlet and triplet
channel are coupled \cite{Gorkov2001,Frigeri2004} and thus the gap
function is a mixture of both channels. However, this coupling is
of the order $\alpha/\epsilon_{F}$ which we take to be small
($\epsilon_F $: the Fermi energy, or analogue to the band width).
Consequently, we ignore this coupling  and consider the singlet
and triplet channels separately.  In particular, it manifests itself in
two ways. It gives rise to different magnitude gaps on the two SOC
spilt Fermi surface sheets. It should also be observable through a
non-vanishing spin-current contribution in the excess current
associated with Andreev scattering.

It has been shown that once $\alpha>>k_BT_{c}$ for the
spin-triplet channel, then the only spin-triplet state that is
permitted satisfies $\vd(\vk)||\vg(\vk)$ \cite{Frigeri2004}. We
assume that $k_BT_c<\alpha<<\epsilon_F$  and solve in sequence the
equations (\ref{Green1}, \ref{Green2}, \ref{Green3}). The
solutions are formally the same for both singlet
($\hat{\Delta}(\vk)= i \{\psi(\vk)\} \hat{\sigma}_{y}$) and
triplet [$\hat{\Delta}(\vk)=i \{\vd(\vk)\cdot \hat{\vsig}\}
\hat{\sigma}_{y}$ with $\vd(\vk)||\vg(\vk)$] pairing. We find
\begin{eqnarray}
\label{regularG}
\hat{G}(\vk,i\omega_n)&=&G_{+}(\vk,\omega_n)\hat{\sigma}_{0}+(\hvg_{\vsk}
\cdot \hat{\vsig}) G_{-}(     \vk,i\omega_n) \nonumber \\
G_{\pm}(i\omega_n)&=&-\frac{1}{2}\left[ \frac{i \omega+E_{+}
  }{(\omega_{n}^{2}+|\Delta|^{2}+E^{2}_{+})} \pm \frac{i \omega+E_{-}
  }{(\omega_{n}^{2}+|\Delta|^{2}+E^{2}_{-})} \right]
\end{eqnarray}
where $E_{\pm}=\xi \pm \alpha |\vg|$, and
\begin{eqnarray}
\label{anomalousF}
\hat{F}(\vk,i\omega_n)&=&[F_+(\vk,i
\omega_n)\hat{\sigma}_{0}+(\hvg_{\vsk} \cdot \hat{\vsig}) F_-(
\vk,i\omega_n)] \hat{\Delta}(\vk), \nonumber \\
\hat{F}^{\dag}(\vk,i\omega_n)&=&\hat{\Delta}^{\dag}(\vk) [F_+(\vk,i
\omega_n)\hat{\sigma}_{0}+(\hvg_{\vsk} \cdot \hat{\vsig}) F_-(
\vk,i\omega_n)] \nonumber\\
       F_\pm(\vk, i\omega_n)&=& \frac{1}{2}\left[ \frac{1
         }{(\omega_{n}^{2}+|\Delta|^{2}+E^{2}_{+})} \pm \frac{1
         }{(\omega_{n}^{2}+|\Delta|^{2}+E^{2}_{-})} \right].
\end{eqnarray}
Note that the anomalous Green's function has less symmetry than
the gap function. In particular, $\Delta(\vk)$ is either symmetric
(singlet) or antisymmetric (triplet) with respect to $ \vk \to - \vk $,
while the resulting anomalous
Green's function has both a symmetric and an antisymmetric
component. This leads to the mixing of the spin-singlet and
spin-triplet Cooper pairs and has been discussed by Gor'kov and
Rashba in the case of 2D metals \cite{Gorkov2001}. However $F_{-}$ is an odd function in $\xi$ and doesn't contribute to the gap equation in its weak coupling formulation. In this way the magnitude of the gap $\psi=\psi(\vk)$ for the singlet s-wave order parameter can be approximated by the standard and universal $BCS$ gap equation,

\begin{eqnarray}
    \label{GapDimLess}
    \ln(\psi)&=&- \int^{\infty}_{-\infty} dx  \frac{1}{\sqrt{\psi^{2}+x^{2}}} \frac{1}{ \exp \left( \frac{\pi}{\gamma}\frac{ \sqrt{\psi^{2}+x^{2}}}{k_{B}T}\right)+1}
\end{eqnarray}

where $T$ is expressed in units of $T_{c}$, $\psi$ in units of $\psi(T=0)$, where $\psi(T=0)/k_{B}T_{c}=\pi/\gamma$  and $C=\ln(\gamma)=0.577$ correspond to the Euler constant. The deviation will be of the order $(\alpha/\epsilon_{c})^{2}$, with $\epsilon_{c}$ the cut-off energy of the attractive interaction, which is small in our case. For the protected spin-triplet state ($\vd(\vk)=\Delta_0 \vg(\vk)$) we have $\Delta_0(T=0) /k_{B}T_{c} \simeq \psi(T=0)/k_{B}T_{c}$, and to avoid numerical complication superfluous for the following discussion, we have set $\Delta_0(T/T_c)= \psi(T/T_c)$.
\section{ \label{Gap} The Static Uniform Spin Susceptibility}
%
In materials with a spatial inversion center, the measurement of
the Knight shift in the resonance frequency ($\delta \omega$) by
nuclear magnetic resonance (NMR) is an important experimental tool
in determining the nature of the superconducting state. In
particular, it allows the determination of the spin structure of
the Cooper pairs. The measurement of  the temperature dependence
of the ratio $\delta \omega_{s}/ \delta \omega_{n}=\chi_{s}/
\chi_{n}$ is a direct measure of the behavior of the  spin
susceptibility of the superconducting state $(\chi_{s})$ ($\chi_n$
is the spin-susceptibility of the normal state). For spin-singlet
superconductors it is known that the paramagnetic susceptibility
is proportional to the density of normal electrons, which vanishes
at zero temperature. In the spin-triplet case, the spin of the
Cooper pairs can contribute to the susceptibility. In particular,
if the external field ($\vH$) is parallel to the spin of the
Cooper pair ($\vH \perp \vd$), then the susceptibility coincides
with that of the Fermi-liquid normal state.  This property was
used to experimentally confirm spin-triplet superconductivity, for
example, in Sr$_{2}$RuO$_{4}$ \cite{Ishida1998}.

In principle, for materials with strong spin-orbit coupling, the
total magnetic susceptibility cannot be split into separate
orbital and spin parts. However, if $\alpha << \epsilon_F$ it is
possible to isolate the two components \cite{Boiko1960}. As shown
in Ref. \cite{Abrikosov1962}, the spin susceptibility tensor
$\chi^{s}_{i j}$ in the superconducting state can be expressed as

\begin{eqnarray}
        \label{}
        \chi^{s}_{i j}&=&-\mu^{2}_{B} k_{B}T \sum_{\vsk} \sum_{\omega_{n}}
        tr \{ \hat{\sigma}_{i} \hat{G}(\vk, \omega_{n})
        \hat{\sigma}_{j} \hat{G}(\vk, \omega_{n}) - \hat{\sigma}_{i}
        \hat{F}(\vk, \omega_{n}) \hat{\sigma}^{\top}_{j}
        \hat{F}^{\dag}(\vk, \omega_{n}) \} \nonumber \\
\end{eqnarray}
If we assume a spherical Fermi surface and a constant density of
states $N(\xi)$ close to the Fermi surface, then we can replace
the sum over $\vk$ by $\sum_{\vk} \to N(0) \int \frac{
d\Omega}{4\pi} \int d\xi$. In the normal state the integral over
$d\xi$ cannot be carried out before the sum over $\omega_{n}$
because the regular Green's function will be formally divergent
\cite{Abrikosov1975}. Doing the sum first, we find that the normal
state spin susceptibility corresponds with the Pauli
susceptibility $\chi_{n}=2\mu_{B}^{2}N(0)$. This would not be the
case if electron-hole asymmetry is taken into account (i.e if
$dN(\xi)/d\xi |_{\xi=0} \neq 0$). For our purposes this is an unnecessary
complication which we will neglect here.

To avoid carrying out the summation over $\omega_n$ in the
superconducting state, we follow Abrikosov et al. and sum and subtract
the expression corresponding to the normal state
\cite{Abrikosov1975}. The integral of the difference between the
integrands rapidly converge in this case. Consequently, the order
of summation and integration can be interchanged. For  a singlet
gap function we find a generalization of the result obtained by
Gork'ov and Rashba \cite{Gorkov2001} and Bulaevskii et al \cite{Bul76}:
\begin{eqnarray}
        \label{SingSusc}
        \chi^{s}_{i i}&=&\chi_{n}
         \left \{ 1-k_{B}T \pi \sum_{\omega_{n}} \left\langle
         \frac{1-\hvg^{2}_{\vsk,i}}{(\omega^{2}_{n}+|\psi(\vk)|^{2}+\alpha^{2}
         |\vg_{\vsk}|^{2})} \cdot
         \frac{|\psi(\vk)|^{2}}{\sqrt{\omega_{n}^{2}+|\psi(\vk)|^{2}}}
         \right. \right. \nonumber \\
         && \qquad \qquad \left. \left. +  \hvg^{2}_{\vsk,i}
         \frac{|\psi(\vk)|^{2}}{(\omega^{2}_{n}+|\psi(\vk)|^{2})^{3/2}}
         \right \rangle_{\vsk} \right\} .
\end{eqnarray}

 For the triplet gap function (with $ \vd(\vk) || \vg_{\vk} $) the
 susceptibility is independent of $\alpha$
\begin{eqnarray}
        \label{TripSusc}
        \chi^{s}_{i i}&=&\chi_{n}
         \left \{ 1-k_{B}T \pi \sum_{\omega_{n}} \left \langle
         \hvg^{2}_{\vsk,i}
         \frac{|\vd(\vk)|^{2}}{(\omega^{2}_{n}+|\vd(\vk)|^{2})^{3/2}}
         \right \rangle_{\vsk} \right\} .
\end{eqnarray}
More precisely, the contribution due to $\alpha$ from the regular
Green's function is cancelled out by the contribution of the
anomalous Green's function.

\begin{figure}[h]
\begin{center}
    \includegraphics[width=6cm, height=5cm ]{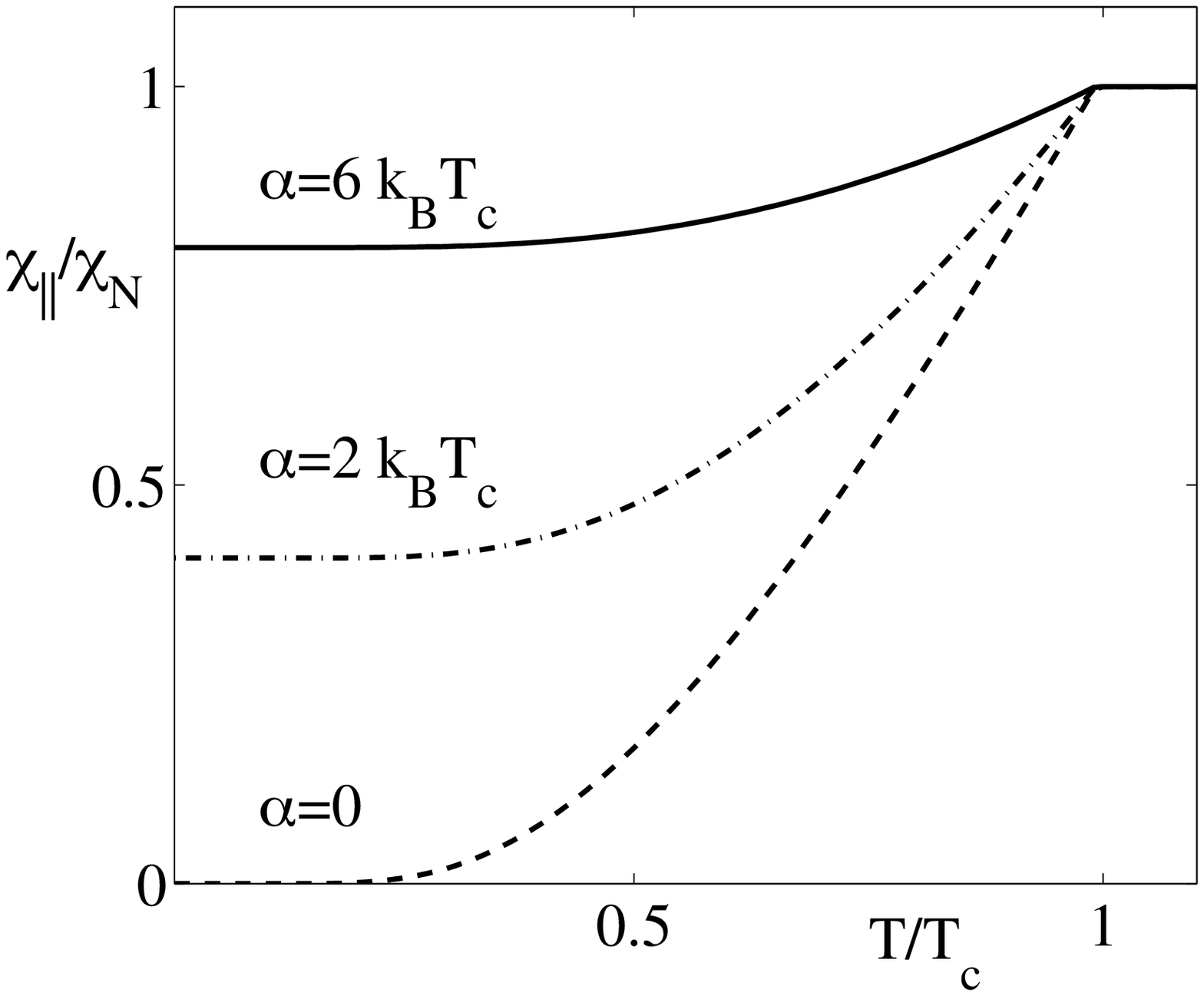}
    \includegraphics[width=6cm, height=5cm ]{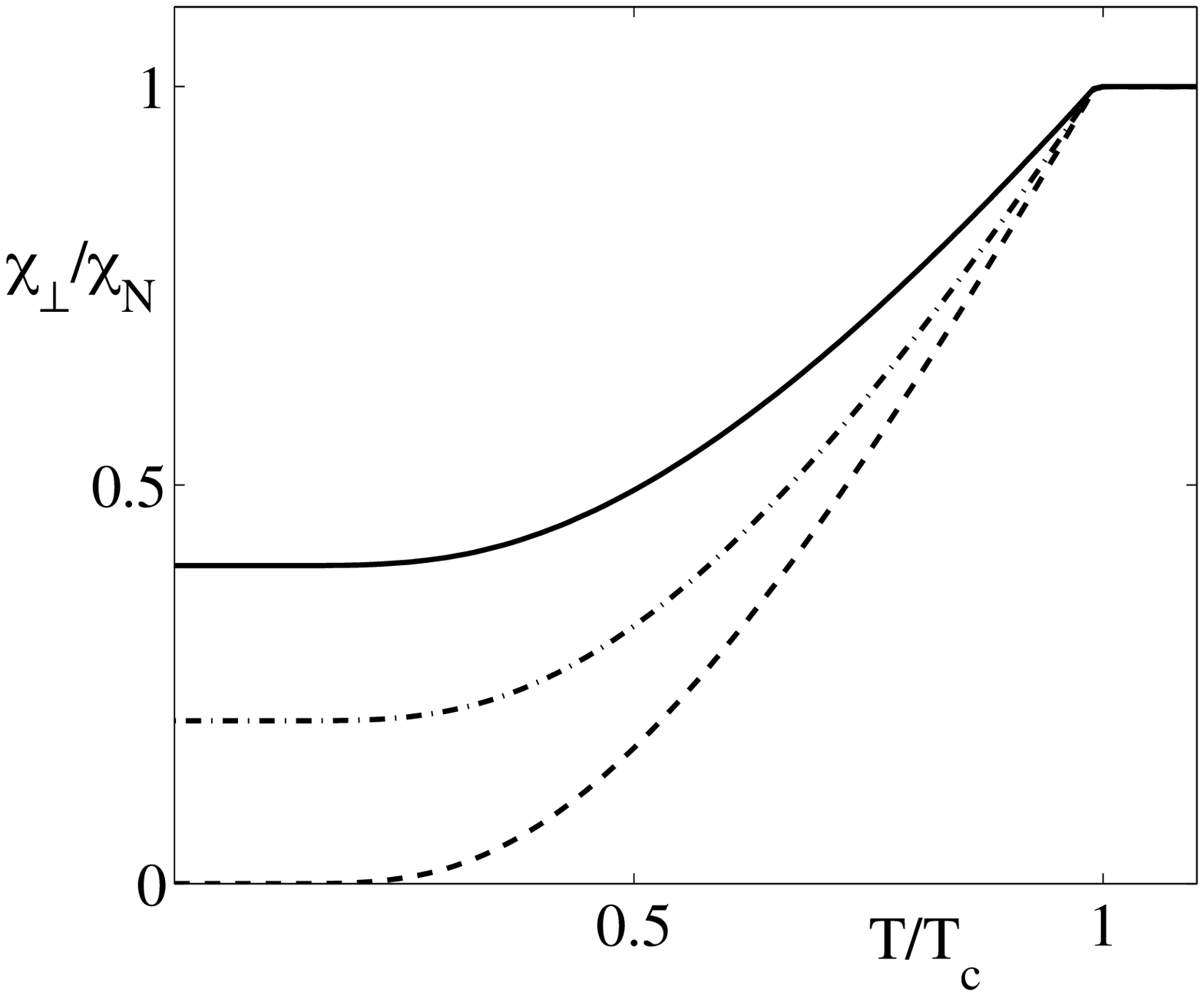}
\end{center}
\caption{\label{SuscT}  The spin susceptibility in case of singlet
    s-wave gap function for $\vg_{\vk} \varpropto (-k_{y},k_{x},0)$
    (CePt$_{3}$Si). The spin susceptibility in the ab-plane
    $\chi_{\perp}$ and along the c-axis $\chi_{\parallel}$ as
    a function of $T$ for three different values of the spin orbit
    coupling $\alpha$.
    The susceptibility in the superconducting state ($T/T_{c}<1$)
increases with the
   spin-orbit coupling strength. The susceptibility is
   more strongly suppressed in the ab-plane than along the c-axis. At $T=0$
   we have $\chi^{s}_{\perp}=\chi^{s}_{\parallel}/2$.}
\end{figure}

\begin{figure}[h]
\begin{center}
    \includegraphics[width=6cm, height=5cm ]{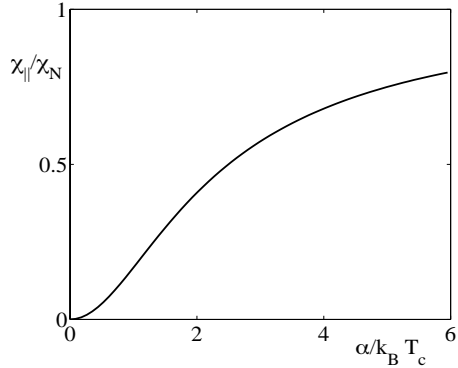}

\end{center}
\caption{\label{Susca} The zero-temperature value of $ \chi^{s}_{\parallel} /\chi_n $ of the spin singlet
state as a function of
$ \alpha $. Note that $ \chi^{s}_{\perp} = \chi^{s}_{\parallel} / 2 $. }
\end{figure}

We now apply the results to the recently discovered heavy fermion
superconductor CePt$_3$Si \cite{Bauer2004}. In this case, the generating
point group symmetry is $C_{4v}$ for which the simplest form of
$\vg_{\vk}$ is $\vg_{\vk}\varpropto \vk \times \mhz =
(-k_{y},k_{x},0)$ \cite{Frigeri2004}. This has the same form as
the well-known Rashba spin-orbit coupling \cite{Rashba1960}. The
spin-susceptibility for the singlet s-wave gap function is shown
in Fig.\ref{SuscT}. The left plot shows the corresponding behavior
of the susceptibility for the field along the c-axis
($\chi_{\parallel}=\chi_{c,c}$). The right plot shows the spin
susceptibility for the field in the ab-plane
($\chi_{\perp}=\chi_{a,a}=\chi_{b,b}$) as function of the
temperature for three different values of the spin-orbit coupling
($\alpha$). In Fig. \ref{Susca} we show how the zero-temperature
value of the susceptibility rises as a function of $ \alpha $. For
$ \alpha \gg k_B T_c $ the $ \chi_{\parallel} $ approaches $ \chi_n $
and $ \chi_{\perp} = \chi_n/2 $.

\begin{figure}[h]
\begin{center}
   \includegraphics[width=6cm, height=5cm ]{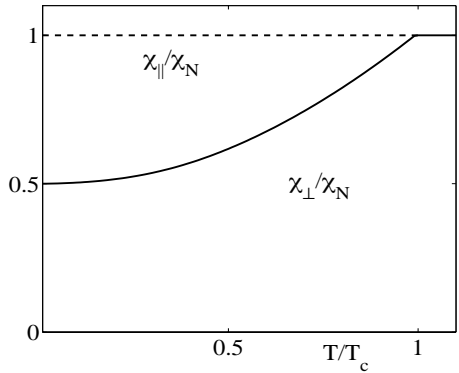}
\end{center}
\caption{\label{SuscTrip}  The spin susceptibility for a
spin-triplet
  p-wave gap function $\vd(\vk) \parallel \vg_{\vk} \varpropto
  (-k_{y},k_{x}, 0)$ (CePt$_{3}$Si). The spin susceptibility in the
  ab-plane  $\chi_{\perp}$ and along the c-axis $\chi_{\parallel}$ as
  function of $T$. The susceptibility is in this case independent of
  the spin orbit coupling $\alpha$. In the superconducting state the
  susceptibility in the ab-plane coincides with that of the normal
  state.}
\end{figure}

We remark that the susceptibility increases with the spin-orbit
coupling strength. For $\alpha$ very large, the resulting
susceptibility looks very similar to that obtained for the triplet
p-wave gap function shown in Fig. \ref{SuscTrip}. For the spin triplet
phase we chose in accordance to Ref.\cite{Frigeri2004} the pairing
state $ \vd = \mhx k_y - \mhy k_x $.
The similar
properties of the spin susceptibilities make it difficult to  distinguish between a spin-triplet
and spin-singlet order parameter through NMR measurements in the
strong SOC limit.

Due the complicated band structure of CePt$_3$Si \cite{Samokhin}
and the coexistence of superconductivity with antiferromagnetism
\cite{metoki2004}, our theory does not provide a quantitative
description for the spin susceptibility for CePt$_3$Si. However,
our approach illustrates the behavior expected at a qualitative
level. More precisely, the susceptibility, independent of the kind
of pairing and of the strength of the SOC, is more strongly
suppressed in the ab-plane than along the c-axis. This angle
dependence of the static uniform susceptibility should be
confirmed by NMR-Knight-shift measurements. Moreover, our
discussion supports the conclusion of Frigeri et al. that the spin
"singlet" pairing state acquires a certain robustness against pair
breaking due to spin polarization \cite{Frigeri2004}. A rough
estimate of the zero-temperature limiting field is obtained by
comparison of superconducting condensation and spin polarization
energy, leading to
\begin{equation}
H_p \approx \frac{k_B T_c}{\mu_B \sqrt{1-\chi^{s}(T=0)/\chi_n}} \;.
\end{equation}
In principle this can become very big for fields along the $c$-axis and roughly 1- 2 Tesla for fields in the
basal plane.
%
\section{Conclusions}

We have determined the spin susceptibility in superconductors
without inversion symmetry. While the spin-triplet and
spin-singlet order parameters are mixed in general, we can discuss
predominantly spin-triplet or spin-singlet gaps when the spin
orbit coupling is much smaller than the band width. We have found
that for the surviving predominantly spin-triplet gap, the lack of
inversion symmetry does not change the spin-susceptibility. For a
predominantly spin-singlet gap, the lack of inversion symmetry
leads to an increase in the spin-susceptibility. For large
spin-orbit coupling (relative to $T_c$), the spin susceptibilities
for both the spin-singlet and spin-triplet gaps become similar.
For the heavy fermion superconductor CePt$_3$Si, we have predicted
that, independent of the pairing symmetry, the susceptibility is
more strongly suppressed in the ab-plane than along the c-axis.
More generally, we may state that the spin susceptibility in a
superconductor without inversion center is approximately described
by the behavior of the spin triplet superconductor with $ \vd
(\vk) = \Delta_0 \vg_{\vsk} $ and is obtained from Eq.
(\ref{TripSusc}).

We are grateful to E. Bauer, R.P Kaur, A. Koga, and T.M. Rice for
many helpful discussions. This work was supported by the Swiss
National Science Foundation. DFA was also supported by the
National Science Foundation grant No. DMR-0381665, the Research
Corporation, and the American Chemical Society Petroleum Research
Fund.


\section*{References}
\begin{thebibliography}{10}
\bibitem{Anderson1959} Anderson P W 1959 {\it J. Phys. Chem. Solids}
  {\bf 11} 26
\bibitem{Anderson1984} Anderson P W 1984 {\it \PR B} {\bf 30} 4000

\bibitem{Bauer2004} Bauer E, Hilscher G, Michor N, Paul Ch, Scheidt E
  W,  Gribanov A, Seropegin Yu, Noeel H, Sigrist M and Rogl P 2004 {\it \PRL} {\bf 92} 027003
\bibitem{C115} Bianchi A, Moshovich R,  Oeschler N,  Gegenwart P, Steglich F, Thompson JD,
Pagliuso PG and Sarrao JL 2002, {\it Phys. Rev. Lett.} {\bf 89} , 137002
  {\it \PRL} {\bf 92} 027003
\bibitem{Frigeri2004} Frigeri P A, Agterberg D F, Koga A, and Sigrist
  M 2004 {\it \PRL} {\bf 92} 097001
\bibitem{Gorkov2001} Gor'kov L P and Rashba E I 2001{\it \PRL} {\bf 87} 0370041
\bibitem{Samokhin}  Samokhin K V, Zijlstra and Bose S 2004 {\it \PR B} {\bf 69} 094514
\bibitem{Dresselhaus1955} Dresselhaus G 1955, Phys. Rev. {\bf 100} 580
\bibitem{agt04} Agterberg DF and Sigrist M, {\it to be published}.
\bibitem{Ishida1998} Ishida  K, Mukuda H, Kitaoka Y, Asayama K, Mao Z
  Q, Mori Y and Maeno Y  1998 {\it Nature} {\bf 396} 658

\bibitem{Boiko1960} Boiko I I and Rashba E I 1960 {\it
    Sov. Phys. Solid State} {\bf 2} 1692
\bibitem{Abrikosov1962} Abrikosov A A and Gor'kov L P 1962 {\it
    Sov. Phys.-JETP} {\bf 15} 752
\bibitem{Abrikosov1975} Abrikosov A  A, Gor'kov L P, and Dzyaloshnskii
  1975 {\it Methods of Quantum Field Theory in Statistical Physics}
  (New York: Dover)
\bibitem{Rashba1960} Rashba E I 1960 {\it Sov. Phys. Solid State} {\bf 2} 1109
\bibitem{metoki2004} Metoki N, Kaneko K, Matsuda T D, Galatanu A, Takeuchi T,
Hashimoto S, Ueda T, Settai R, Onuki Y, and N Bernhoeft N 2004
{\it J. Phys.: Condens. Matter} {\bf 16} L207
  \bibitem{Bul76} Bulaevskii L N, Guseinov A A and Rusinov A I 1976, Sov. Phys. JETP {\bf 44} 1243
\end{thebibliography}
\end{document}